\documentclass[12pt]{article}
\usepackage{epsfig,graphicx}
\usepackage{fpcp03}

\newcommand{\lsime} {\buildrel < \over {_\sim}}
\newcommand{\gsime} {\buildrel > \over {_\sim}}
\begin{document}


\Title{Phenomenology of New Physics 
\footnote{Invited talk at FPCP
2003, the 2nd Conference on Flavor Physics And CP Violation,
3-6 June 2003, Paris, France; to be published in the proceedings.}
}

\begin{flushright}
LMU 17/03\\
August 2003\\
\end{flushright}
\vspace{-2cm}


%
\label{HillerStart}

%
\author{ Gudrun Hiller\index{Hiller, G.}}

%
\address{Ludwig-Maximilians-Universit\"at M\"unchen, Sektion Physik \\
Theresienstra\ss{}e 37, D-80333 M\"unchen, Germany\\
}

\makeauthor\abstracts{
I review the phenomenology of new physics in low energy processes 
using the notion of minimal flavor violation (vs.~non-minimal
flavor violation).
I compare the predictions of beyond-the-standard models and show that 
among certain observables in rare $b$-decays pattern arise, which allow to 
distinguish between extensions of the Standard Model.
I discuss the status and future 
of the model independent analysis of $b \to s$ processes.
}

\section{Introduction}

There are several reasons why we are unhappy with the Standard Model
(SM).
We observed phenomena which are not part of the SM, such as finite
neutrino masses, dark energy $ \Omega_{DE} \simeq 75 \%$, gravity and the
matter anti-matter asymmetry $(n-\bar n)/s \sim 10^{-10}$.
We do have questions which cannot be answered within the SM. For
example,
about unification and the origin of flavor and breaking
of CP symmetry because in the SM 
the CKM matrix elements and fermion masses (also in the lepton sector)
are just parameters.
Moreover, the SM has consistency problems.
There is the strong CP problem, i.e.~why is the CKM phase order one 
whereas the strong phase is small $\bar \theta \leq 10^{-10}$
and the gauge hierarchy problem.
In the SM, scalar masses receive quadratic radiative corrections 
$\delta m^2 \sim \Lambda^2/16 \pi^2$.
For a high cut-off such as the Planck scale $\Lambda \sim \Lambda_{Pl}$
a huge amount of fine-tuning is required
to render the renormalized scalar, i.e.~Higgs mass of the order of
the electroweak scale. In other words,
the SM is only natural up to $\Lambda \sim 1$ TeV. Excitingly,
we probe even higher energies in the near future at the Tevatron and the LHC.

\begin{figure}[htb]
\begin{center}
\epsfig{file=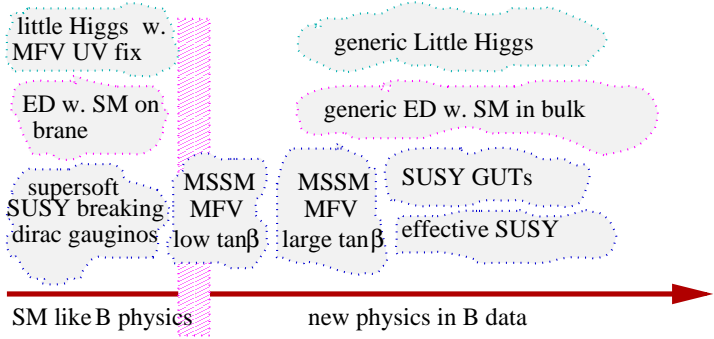,height=45mm}
\caption{Flavor/CP yield of models of electroweak symmetry breaking.
Figure taken from \cite{Hiller:2002cb}.}
\label{fig:hiller-fig1}
\end{center}
\end{figure}

Models of electroweak symmetry breaking where the Higgs masses are
protected can be build by using
supersymmetry (SUSY), extra dimensions, strong
dynamics (technicolor,little Higgs theories) plus hybrids.
In all of these extensions of the SM we expect to see new physics 
(NP) at the TeV scale.
The reach in indirect signals below 5 GeV 
such as in rare $b,c,K,\tau$ decays, meson
mixing and electric dipole moments depends sensitively on how much
beyond the SM flavor and/or CP violation is in the model.
This is illustrated in Fig.~\ref{fig:hiller-fig1}, where the 
prospects for NP in $b$-data are shown as a function of a particular
realization of a model type, for details see \cite{Hiller:2002cb}.

It is customary to classify NP models into those which are
{\it minimal flavor violating} (MFV) and those who are not.
A model is MFV if it does not contain more flavor and CP violation
than the SM, i.e.~what is contained in the Yukawas (CKM). We come back
to a formal definition in Section 
\ref{sec:hiller-modelindependent}.
As an example, the two Higgs doublet models (2HDM) I and II are MFV.
The same is true for the
minimal supersymmetric standard model (MSSM) with flavor blind SUSY
breaking and no further CP violation such as gauge mediated SUSY
breaking with $A$-terms being proportional to the corresponding
Yukawas and squark masses proportional to the unit matrix.
(We neglect small effects from renormalization group running.)
Non-MFV models are the 2HDM III with tree level flavor changing neutral
currents (FCNC), models with fourth
generation quarks, vector like down quarks with tree level FCNC to the
$Z$ and the generic MSSM with/or without R-parity conservation.
Hence, MFV theories require very different model building from those
which are not.

Experimental signals for non-MFV include
{\it i} non-standard CP violation, 
e.g.~$\sin 2 \beta(\phi K_S) \neq \sin 2 \beta( J/\Psi K_S) $,
{\it ii}
right-handed currents, which are generically suppressed in $b \to s $ 
transitions in MFV models $\sim m_s/m_b$,
\begin{figure}[htb]
\begin{center}
\epsfig{file=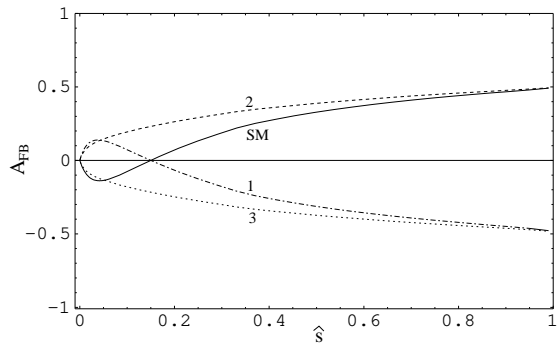,height=45mm}
\caption{Shapes of the Forward-Backward asymmetry in 
$b \to s \ell^+ \ell^-$ decays in the SM (solid) and three beyond the
SM scenarios.
The curves 1 and 3 (or a flat asymmetry $A_{FB}(\hat s) \sim 0$) 
require non-MFV.
Figure taken from \cite{Ali:2002jg}.}
\label{fig:hiller-afb}
\end{center}
\end{figure}
{\it iii} certain 
shapes of  the Forward-Backward asymmetry  $A_{FB}$ for inclusive and 
exclusive $B \to (X_s,K^*) \ell^+ \ell^-$ decays, see
Fig.~\ref{fig:hiller-afb}, where currently allowed possibilities are shown.
Note that the displayed curves exhibit discrete differences rather
than being gradually distinct, hence, can be cleanly investigated with
the exclusive decay \cite{Ali:1999mm}.
Furthermore, {\it iv} beyond MFV there is
no ``CKM-link'' among $b \to s, b \to d$ and  $s \to d$ transitions.

There is an existent $2.7 \sigma$ 
hint for NP which is non-MFV, namely beyond the SM CP violation in $B \to
\phi K_S$ decay, see Table \ref{tab:hiller-phiKs}
\footnote{During the completion of this 
write-up both BaBar and Belle issued improved measurements of
$S_{\phi K_S}^{BaBar}=+0.45 \pm0.43 \pm 0.07$,
$C_{\phi K_S}^{BaBar}=-0.38 \pm0.37 \pm0.12$
and
$S_{\phi K_S}^{Belle}=-0.96 \pm0.50^{+0.09}_{-0.11}$,
$C_{\phi K_S}^{Belle}=+0.15 \pm 0.29 \pm 0.07$
based on larger data samples \cite{browderLP03}. The
new error weighted averages are
$S_{\phi K_S}^{ave}=-0.15 \pm 0.33$, still 2.7 $\sigma$
away from the SM and 
$C_{\phi K_S}^{ave}=-0.05 \pm 0.24$, i.e.~consistent with 
small direct CP violation.
We note, however, that there is no good agreement between the two
experiments in $S_{\phi K_S}$.
Note also the new Belle result 
$S_{\eta^\prime K_S}^{Belle}=+0.43 \pm0.27 \pm 0.05$ updating
$S_{\eta^\prime K_S}^{Belle2002}=+0.71 \pm0.37^{+0.05}_{-0.06}$, which 
with $S_{\eta^\prime K_S}^{BaBar}=+0.02 \pm0.34 \pm 0.03$ leads to
$S_{\eta^\prime K_S}^{ave}=+0.27 \pm 0.21$ \cite{browderLP03}.}.
Data yield $\sin 2 \beta_{ave}=+0.736 \pm 0.049$ \cite{browderLP03},
in agreement with the fit to the unitarity triangle 
$\sin 2 \beta_{UT fit}=+0.74 \pm 0.10$ @95\%C.L.~\cite{CKM-fit} and 
$\lambda \simeq 0.22$.
\begin{table}[htbp]
\renewcommand{\arraystretch}{1.0}
         \begin{center}
         \begin{tabular}{|c|c|c|c|c|}
\hline
\mbox{} & BaBar \cite{babarphi}& Belle \cite{bellephi}& average & SM+MFV \\
\hline
$S_{\phi K_S}$ &$ -0.18 \pm 0.51 \pm 0.07 $&$ -0.73{\pm 0.64}\pm 0.22$
&  $ -0.38\pm 0.41$ & $\sin 2 \beta + {\cal{O}}(\lambda^2)$ \\
$C_{\phi K_S}$ &$ -0.80 \pm 0.38 \pm 0.12$ & $+0.56 \pm 0.41 \pm 0.16$
& $-0.19 \pm 0.30 $& ${\cal{O}}(\lambda^2)$ \\
\hline
         \end{tabular}
\caption{Data  on 
time dependent asymmetries in $B \to \phi K_S$ vs.~SM and MFV theories.} 
\label{tab:hiller-phiKs}
         \end{center}
\end{table}
The ${\cal{O}}(\lambda^2)$ correction 
from the $u \bar u$ loop maybe dynamically enhanced
\cite{Grossman:1997gr}, e.g.~by large rescattering.
This  SM background can be constrained using
SU(3) flavor analysis. 
Currently, we have the not very stringent bound
$|\xi_{\phi K^0}| \leq 0.25$, where
$|\sin 2 \beta (\phi K_S)-\sin 2 \beta | \leq 2 \cos 2 \beta |\xi_{\phi K^0}|$
\cite{Grossman:2003qp}. It is derived from upper bounds on 
${\cal{B}}(B^+ \to \phi \pi^+)$, ${\cal{B}}(B^+ \to \bar K^{*0} K^+)$ and
can be experimentally improved soon.
It assumes that no large amplitudes in the charged $B$-decay cancel,
i.e.~$|\xi_{\phi K^0}| \leq |\xi_{\phi K^+}|$. 
The bound can be made independent of this
assumption by improved data on 11 further branching
ratios, see \cite{Grossman:2003qp}.
This will be important
if experimental errors on $S_{\phi K_S}$ shrink and the 
central value moves closer to the SM expectation.
Note that one obtains  $|\xi_{\eta^\prime K^0}|\leq 0.36 $ or
$|\xi_{\eta^\prime K^0}|$ $\leq$  $|\xi_{\eta^\prime K^+}| \leq 0.09 $, 
if $N_c$ counting works for the tree level 
contributions to $B^{0/\pm} \to \eta^\prime K^{0/\pm}$
\cite{Grossman:2003qp}.
 
\section{Models with non-MFV}

In order to obtain the current central value of $S_{\phi K_S}$ an 
$O(1)$ NP contribution
with an $O(1)$ CP phase is required on the decay amplitude 
\cite{Hiller:2002ci,Atwood:2003tg}.
This NP can be in the coefficients of 
QCD $C_{3,\ldots 6}^{(\prime)}$, electroweak penguins 
$C_{7,\ldots 10}^{(\prime)}$
and/or the  chromomagnetic dipole $ C_{8 g}^{(\prime)}$ \cite{Kagan:1997sg}.
(The operators are e.g.~given in \cite{Atwood:2003tg}.)
We discuss two possible explanations and show how to distinguish them.

Non-SM $sZb$-couplings arise generically
in many models such as vector like down quarks, 4th generation,
non-MFV SUSY, anomalous couplings, $Z^\prime$ models.
They can be written as
\begin{eqnarray}
{\cal{L}}_{Z}=\frac{g^2}{4 \pi^2} \frac{g}{2 \cos \Theta_W}
(\bar b_L \gamma_\mu s_L Z_{sb} +\bar b_R \gamma_\mu s_R Z^\prime_{sb})Z^\mu 
+h.c.
\end{eqnarray}
They modify the coefficients of the 4-Fermi operators 
$O_{3,7,9}^{(\prime)}$ 
which contribute to $b \to s \bar s s$ decays \cite{Atwood:2003tg}.
The $sZb$-couplings are experimentally constrained as 
\begin{eqnarray}
\label{eq:bound}
\sqrt{|Z_{sb}+Z_{sb}^{SM}|^2 +|Z_{sb}^\prime|^2} \leq 0.08 ~~~~~ 
Z_{sb}^{SM*}=-V_{tb}V_{ts}^* \sin^2 \Theta_W C_{10 \ell}^{SM} \simeq
-0.04
\end{eqnarray}
The bound in Eq.~(\ref{eq:bound}) is based on inclusive $B \to X_s e^+ e^-$
decays at NNLO \cite{Ali:2002jg}
and corresponds to an enhancement of 
2 to 3  over the SM value. 
$Z_{sb}^{(\prime)}$ large and complex can explain the anomaly in 
$B \to \phi K_S$ decay \cite{Hiller:2002ci}.
The implications of anomalous $sZb$-couplings include distortion of
dilepton sprectra and the $A_{FB}$ shape in $b \to s \ell^+ \ell^-$
decays and the $b \to s \nu \bar \nu$ branching ratio. They further 
induce a non-zero Forward-Backward-CP asymmetry 
$A_{FB}^{CP}  \equiv  \frac{A_{FB}+\bar{A}_{FB}}{A_{FB}-\bar{A}_{FB}}  
\sim \frac{Im(C_{10 \ell})}{Re(C_{10 \ell})}$ which probes the phase of the
$s Z b$ vertex. The SM background is tiny $A_{FB}^{CP}< 10^{-3}$ 
\cite{Buchalla:2000sk}. 
There is experimental support for the possibility of 
large electroweak penguins in $B \to K \pi$ decays \cite{Yoshikawa:2003hb},
which, for example, could be induced by non-standard $Z$-penguins.
\begin{table}[htbp]
\renewcommand{\arraystretch}{1.0}
 \begin{center}
\begin{tabular}{|c|c|c|}
\hline
\mbox{} & $Z$-penguins &  MSSM with $(\delta^D_{23})_{RR}$ \\
\hline
${\cal{B}}(b \! \to \! s \ell^+ \ell^- \!),A_{FB}(b \! \to \! s \ell^+ \ell^- \!)$ & up to ${\cal{O}}(1)$ 
effects 
& MFV MSSM like \cite{Ali:2002jg} \\
${\cal{B}}(B_s \to \mu^+ \mu^-)$ & up to ${\cal{O}}(10) \cdot
{\cal{B}}_{SM}$ \cite{Buchalla:2000sk}
& up to $ {\cal{B}}_{exp. \; bound} \sim {\cal{O}}(10^3) \cdot {\cal{B}}_{SM}$ \\
$\Delta m_s$ & $ \mbox{up to} \;  0.5 \cdot \Delta m_{s \; SM}$
\cite{Atwood:2003tg} & $\approx \! \Delta m_{s \; SM}$ up to few
$\! 100 \; ps^{-1}$ \\
$b \to s \gamma$ helicity flip& SM like  
& $|C_{7 \gamma}(\mu_b)^\prime/C_{7 \gamma}(\mu_b)| \lsime 0.4 $  \\
$a_{CP}(b \to s \gamma)$ & SM like & SM like    \\
\hline
\end{tabular}
\caption{Predictions of two beyond the SM models.} 
\label{tab:hiller-corr}
\end{center}
\end{table}

The MSSM with large and complex mixing between right-handed 
$\tilde s$ and $\tilde b$, denoted here as $(\delta^D_{23})_{RR}$
(which is inspired from large $\nu_\mu-\nu_\tau$ mixing in $SO(10)$ GUTs)
can accommodate large departures in
$S_{\phi K_S}$ from the SM \cite{Harnik:2002vs},
for other recent studies of gluino  mediated
effects in $B \to \phi K_S$ decay see \cite{other}.
The model gives contributions to the flipped 
4-Fermi $O_{3 \ldots 6}^\prime$ and dipole operators 
$O_{7 \gamma}^\prime$, $O_{8 g}^\prime$. 
An enhancement of ${\cal{B}}(b \to s \gamma)$ can be avoided by
having the gluino mass sufficiently lighter than the squark masses.
The $B_s$-$\bar B_s$ mixing can be huge 
$\Delta m_s \sim 100 \; ps^{-1}$. The presence of
large right handed currents imply 
flipped helicity contributions to $b \to s \gamma$, 
see Section \ref{sec:hiller-modelindependent}.
Direct CP violation in $b \to s \gamma$ is SM like, since
only flipped coefficients have a NP phase and
different helicities do not interfere.

\begin{figure}[htb]
\begin{center}
\epsfig{file=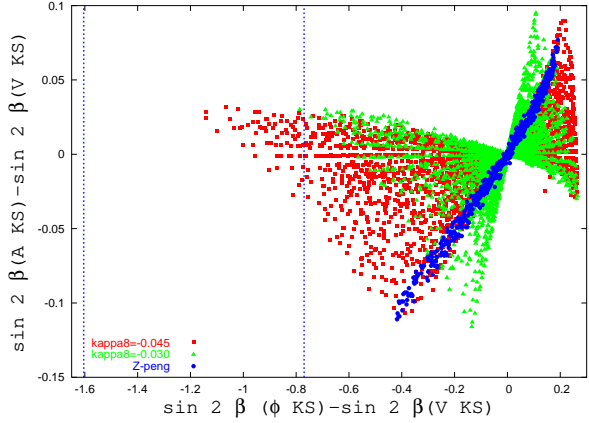,height=80mm,angle=270}
\caption{$ \sin 2\beta_{A K_S}-
\sin 2\beta_{V K_S}$ as a function of
$\sin 2\beta_{\phi K_S}-\sin 2\beta_{V K_S}$
in the non-SM $Z$-scenario (blue) and in the MSSM with additional 
flavor violation induced by $\delta^D_{RR \; 23}$.
The latter butterfly type correlation is shown for two values of
the matrix element of
${\cal{O}}_{8 g}^{(\prime)}$. Figure taken from \cite{Atwood:2003tg}.}
\label{fig:hiller-ccbar}
\end{center}
\end{figure}

Predictions of both non-MFV models
are compared in Table \ref{tab:hiller-corr}.
Further means to distinguish them is  to study 
CP asymmetries of the ``golden'' modes $B \to (c \bar c) K $ 
\cite{Atwood:2003tg}. 
Order one NP in $b \to s \bar s s$ decays implies 
${\cal{O}}( 10 \%)$ effects in $b \to c \bar c s$, which is within 
the errors of the UT fit. 
Since the  NP effect is split among final states with the 
same flavor content but different CP quantum numbers we compare vector
$V=J/\Psi,\Psi^\prime$ and axial vector
$A=\chi_1,\eta_c$ coupling charmonia. Current data
$\sin \!2 \beta (A K_S) \! -\sin 2 \beta (V K_S)= \!-0.05 \pm 0.26$
\cite{Aubert:2002ic}
are not significant yet. 
The correlation with $\phi K_S$ is shown in
Fig.~\ref{fig:hiller-ccbar} for both models.
Since
$C_{8 g}^{(\prime)}$ is color octet suppressed in $b\to c \bar c s$
decays this distinguishes NP in the chromomagnetic dipole 
from NP in the 4-Fermi operators.
Note that this SM test is 
independent of improvement of the UT fit.

\section{Model independent analysis \label{sec:hiller-modelindependent}}

The search for NP in $b \to s$ transitions can systematically be
performed in terms of an effective low energy theory
${\mathcal{H}}_{eff}=-4 G_F / \sqrt{2} V_{tb} V_{ts}^* \sum_i
(C_i {\mathcal{O}}_{i} + C_i^\prime {\mathcal{O}}_{i}^\prime )$.
Important operators are given in Table \ref{tab:hiller-opertors}.
The primed (flipped) operators are obtained from interchanging
$L \leftrightarrow R$.
The coefficients of the SM operator basis 
$C_{7 \gamma},C_{8 g},C_{9 \ell},C_{10 \ell}$ have been studied 
recently \cite{Ali:2002jg}.
{}From the $b \to s \gamma$ branching ratio bounds in the 
$C_{7 \gamma}$-$C_{8 g}$ plane have been obtained, allowing for two
different solutions with different sign of $C_{7 \gamma}$ each of which
can be accessed in the MFV MSSM.
There is a bound $|C_{8 g}(m_W)/C_{8 g \, SM}(m_W)| \leq 10 $ 
from charmless $B$-decay and theory input \cite{Greub:2000sy}.
Constraints on the dilepton couplings $C_{9 \ell}$-$C_{10 \ell}$
for each branch have been worked out from $b \to s \ell^+ \ell^-$ decays.
Currently, the inclusive $B \to X_s \ell^+ \ell^-$
(with $\sqrt{q^2} > 0.2$ GeV) and 
exclusive $B \to K \ell^+ \ell^-$ decays 
with electron and muon modes combined ($\ell=e,\mu$) have been observed
\footnote{New numbers were presented at LP03 \cite{NakaoLP03}, 
i.e.~${\mathcal{B}}(B \to K \ell^+ \ell^-)_{BaBar} = (0.69^{+0.15}_{-0.13}
\pm 0.06) \cdot 10^{-6}$,
${\mathcal{B}}(B \to K \ell^+ \ell^-)_{Belle}  =  
(0.48^{+0.10}_{-0.09} \pm 0.03 \pm0.01) \cdot 10^{-6}$
and Belle`s observation of the $K^*$ mode
${\mathcal{B}}(B \to K^* \ell^+ \ell^-)_{Belle}  =  
(1.15^{+0.26}_{-0.24} \pm 0.07 \pm0.04) \cdot 10^{-6}$.}
\begin{eqnarray}
{\mathcal{B}}(B \to X_s \ell^+ \ell^-) & = &
(6.1 \pm 1.4^{+1.4}_{-1.1}) \cdot 10^{-6} \;\;\;\;\;\;\;\; (Belle 
\cite{Kaneko:2002mr}) \\
{\mathcal{B}}(B \to X_s \ell^+ \ell^-) & = &
(6.3 \pm 1.6^{+1.8}_{-1.5}) \cdot 10^{-6} \;\;\;\;\;\;\;\; 
(BaBar \cite{Aubert:2003rv}) \\
{\mathcal{B}}(B \to K \ell^+ \ell^-) & = & 
(0.78^{+0.24+0.11}_{-0.20-0.18}) \cdot 10^{-6} 
\;\;\;\;\;\;\;\; (BaBar\cite{Aubert:2002pj}) \\
{\mathcal{B}}(B \to K \ell^+ \ell^-) & = 
&( 0.58^{+0.17}_{-0.15} \pm 0.06 ) \cdot 10^{-6} 
\;\;\;\; (Belle\cite{belleICHEP}) 
\end{eqnarray}
They are in agreement with the SM
${\mathcal{B}}(B \to X_s \ell^+ \ell^-)_{SM}=4.2 \pm 0.7 \cdot 10^{-6} $ and 
${\mathcal{B}}(B \to K \ell^+ \ell^-)_{SM}= 0.35 \pm 0.12 \cdot 10^{-6}$
for the same cuts \cite{Ali:2002jg}.
While the use of ${\cal{B}}(b \to s \gamma)$ here
is without further
progress currently exhausted by theory 
errors, semileptonic rare decays will yield much information
in the near future beyond branching ratios, in particular from
$A_{FB}$.
Note that curve 2 in Fig.~\ref{fig:hiller-afb} corresponds to the non-SM sign
solution to $C_{7 \gamma}$. The  $A_{FB}$ has a zero in the SM, 
see Fig.~\ref{fig:hiller-afb}, which position is known to high accuracy
for inclusive $\hat s_{SM}^{NNLL}=0.162 \pm 0.002(8)$ 
\cite{Ghinculov:2002pe}
and exclusive $B \to K^* \ell^+ \ell^-$ decays
\cite{Ali:1999mm,Burdman:1995ks} 
$q_{0 \, SM}^{2 \,NNLL}=4.2 \pm 0.6 \mbox{GeV}^2$
\cite{Beneke:2001at}. 
 
\begin{table}[ht]
\renewcommand{\arraystretch}{1.0}
 \begin{center}
\begin{tabular}{|c|c|c|c|}
\hline
operator & magnitude &  phase &  helicity flip ${\cal{O}}_i^\prime$\\
\hline
${\cal{O}}_{7 \gamma} \sim m_b \bar s_L \sigma_{\mu \nu} b_R F^{\mu \nu} $ &
$ b \to s \gamma $ & $a_{CP} (b \to s \gamma)$ & 
$\Lambda_b \to \Lambda \gamma $ 
\\
 & & & $B \to \! (K^*\! \to K \pi) \ell^+ \ell^-$ \\
 & & & $B \to \! (K^{**}\! \to K \pi \pi) \gamma$ \\
\hline
${\cal{O}}_{8 g} \sim m_b \bar s_{L \alpha} 
\sigma_{\mu \nu}T^a_{\alpha \beta} b_{R \beta} G^{a\mu \nu} $ &
$ b \to s \gamma $ & $a_{CP} (b \to s \gamma)$ & $\Lambda_b \to \Lambda \phi $
\\
 & $B \to \not\!\!{X}_c $ & $B \to K \phi$  & $B \to K^* \phi$ \\
\hline
${\cal{O}}_{9 \ell (10 \ell)} \! \sim \!\bar s_L \gamma_\mu b_L \bar \ell \gamma^\mu(\gamma_5) \ell $ & $ b \to s e^+ e^- $ & $A_{FB}(b \to s  \ell^+ \ell^-)$ &
$B \to \!(K^*\! \to K \pi) \ell^+ \ell^-$ \\
\hline
${\cal{O}}_{S(P)} \sim \bar s_L b_R \bar \ell (\gamma_5) \ell $ &
$ B_{d,s} \to \mu^+ \mu^-$ & $B_{d,s} \to \tau^+ \tau^- $
& $b \to s \tau^+ \tau^- $ \\
\hline
\end{tabular}
\caption{FCNC vertices and where they can be tested.} 
\label{tab:hiller-opertors}
\end{center}
\end{table}

In the SM  the scalar/pseudoscalar couplings
$C_{S,P}^{SM} \sim m_\ell m_b/m_W^2$ are very small even for 
$\ell=\tau$, but they can be important in the MFV MSSM 
at large $\tan \beta$.
Constraints on $C_{S,P}$ from 
$B_s \to \mu^+ \mu^-$ decay have been worked out
\cite{Bobeth:2001sq}.
This decay is helicity suppressed in 
the SM with ${\cal{B}}(B_s \to \mu^+ \mu^-)_{SM}=3.2 \pm 1.5 \times
10^{-9}$ like
the corresponding $B_d$-decay 
${\cal{B}}(B_d \to \mu^+ \mu^-)_{SM}={\cal{O}}(10^{-10})$
\cite{Bobeth:2002ch}.
Substantially smaller errors can be obtained using the correlation
(even in some models beyond the SM) with 
the measured values of  $\Delta m_{d,s}$ thus getting rid of the decay
constant and CKM
dependence of the $B_{d,s} \to \mu^+ \mu^-$ branching ratio 
\cite{Buras:2003td}.
Current upper 90 \% C.L.~bounds are
${\cal{B}}(B_d \to \mu^+ \mu^-)<1.6 \cdot 10^{-7}$ (Belle)
and 
${\cal{B}}(B_s \to \mu^+ \mu^-)<9.5 \cdot 10^{-7}$ (CDF Run II)
\cite{NakaoLP03}.
The MFV MSSM predicts interesting correlations, namely 
barring large cancellations that
${\cal{B}}(B_{d,s} \to \mu^+ \mu^-)$ and $\Delta m_s$ 
cannot be both enhanced w.r.t.~their SM values \cite{Buras:2002wq} 
and that the ratio 
${\cal{B}}(B_{d} \to \mu^+ \mu^-)/{\cal{B}}(B_{s} \to \mu^+ \mu^-)\simeq
|V_{td}/V_{ts}|^2$ holds.
The latter can be broken by $O(1)$ beyond minimal models \cite{Bobeth:2002ch}.
Note that $\Delta m_d/\Delta m_s$ does not follow this
pattern of CKM hierarchy in the MFV MSSM \cite{Buras:2001mb}.

So far only a small fraction of Table \ref{tab:hiller-opertors}
has been experimentally accessed.
This program can be extended by allowing for CP phases
\cite{Kagan:1998bh}, taking more than the SM operators into account 
\cite{HK03}, search for right handed currents, e.g.~with polarization 
studies in $\Lambda_b$ \cite{Mannel:1997pc}, radiative $B$
\cite{Gronau:2001ng}
and $B \to (K^* \to K \pi) \ell^+\ell^-$ decays \cite{Kruger:1999xa}. 
Hadronic $b$-decays are sensitive to NP in
Four-quark operators and ${\cal{O}}_{8 g}^{(\prime)}$, however, their
interpretation in terms of the $C_i^{(\prime)}$ 
suffers from hadronic uncertainties.
In some cases it is possible to identify classes of operators, 
e.g.~\cite{Atwood:2003tg,Yoshikawa:2003hb}.

The term MFV can be defined within an effective field theory 
picture. Let the SM be valid  up to a cut-off $\Lambda$, the scale of NP
${\cal{L}}={\cal{L}}_{SM}+ \sum_i {\cal{O}}_i^{(n)}/\Lambda^n $
\cite{Chivukula:1987py,D'Ambrosio:2002ex}.
The gauge sector of the SM, i.e.~${\cal{L}}_{SM}$ with all the 
Yukawas switched off
$Y_u=Y_d=Y_\ell=0$ possesses a $G_F=U(3)^5$ flavor symmetry.
Postulate now that $G_F$ is exact but only broken by the Yukawas
which are interpreted as fields which get a vev $Y \simeq < \phi >$.
Then the effective theory is called MFV if all operators ${\cal{O}}_i^{(n)}$
constructed from the SM and the ``$Y$'' fields are invariant under $G_F$.
Phenomenological bounds from meson mixing and rare decays give
(with 1 Higgs doublet) 
$\Lambda \gsime \mbox{few}$ TeV, similar to the ones from
electroweak precision data \cite{D'Ambrosio:2002ex}.
If nature turns out to be of the MFV kind, this might be an appropriate
model independent frame work also with strong couplings at $\Lambda$.

\section{Summary}

With NP @ TeV the impact on low energy observables depends on the 
amount of 
flavor/CP violation, the presence of large parameters 
(e.g.~$\tan \beta$ in models with two Higgs doublets), 
the actual new particle spectrum and errors.
Order one signals are possible in $b \to s$ processes beyond
MFV, e.g.~in 
$A_{FB}(B \to (X_s,K^*) \ell^+ \ell^-)$.
This is very complementary to direct collider searches which probe the
flavor diagonal sector of the theory. 
``SM-zero'' observables might return surprises,
such as 
searches for non-SM helicity operators  or
differences in CP asymmetries $\sin 2 \beta( c \bar c_A K) -
\sin 2 \beta( c \bar c_V K)$.
While non-MFV models do have a richer phenomenology in rare processes,
there can be sizeable effects in MFV ones as well.
For example, large MFV contributions to the helicity flip operators
$\bar q^\prime_L \Gamma b_R$ in the MSSM at large $\tan \beta$  
lead to an enhanced ${\cal{B}}(B_{d,s} \to \mu^+ \mu^-)$ which
at the same time strongly favors $\Delta m_s$ to be below its SM value.
A precision long term study in semileptonic FCNC`s 
$b \to s \ell^+ \ell^-, s \nu \bar \nu$,
$s \to d \ell^+ \ell^-, d \nu \bar \nu$ decays is promising to test the
SM within a
potential MFV paradigm.
Currently the most salient indication for non-MFV physics beyond the
SM is in $B \to \phi K_S$ decays.
Further experimental study of rare processes will decide
whether MFV is realized or not hopefully soon.

%
\label{HillerEnd}
 
\end{document}